\documentclass{aa}

\usepackage{txfonts}
\usepackage{supertabular}
\usepackage{rotating}
\usepackage{dsfont}
\usepackage{graphicx}
\usepackage{color}
\usepackage{tabularx}
\usepackage{calc}
\usepackage{array}
\usepackage[authoryear]{natbib}

\begin{document}

\titlerunning{Quadrant asymmetry in the angular distribution of the CMB}

\title{Quadrant asymmetry in the angular distribution of cosmic microwave background in the Planck satellite data}

\author{L. Santos \inst{1},  P. Cabella \inst{1,2}, T. Villela \inst{3},  A. Balbi \inst{1,2},  N. Vittorio \inst{1,2} and  C. A. Wuensche  \inst{3} }

\institute{Universit\`a di Roma ``Tor Vergata'', Dipartimento di Fisica, Rome, Italy \and INFN, Sezione di Roma Tor Vergata, Rome, Italy \and Instituto Nacional de Pesquisas Espaciais - INPE, Divis\~ao de Astrof\'isica, S\~ao Jos\'e dos Campos, SP, Brazil}

\offprints{L. Santos \email{ larissa.santos@roma2.infn.it}}

\date{Received      /Accepted}

\abstract  {Some peculiar features found in the angular distribution of the cosmic microwave background (CMB) measured by the Wilkinson Microwave Anisotropy Probe (WMAP) deserve further investigation. Among these peculiar features, is the quadrant asymmetry, which is likely related to the north-south asymmetry.}  
{In this paper, we aim to extend the analysis of the quadrant asymmetry in the $\Lambda$CDM framework to the Planck  foreground-cleaned maps, using the mask provided by Planck team.}  
{We compute the two-point correlation function (TPCF) from each quadrant of the Planck CMB sky maps, and compare the result with 1000 Monte Carlo (MC) simulations generated assuming the $\Lambda$CDM best-fit power spectrum.} 
{ We detect the presence of an excess of power in the southeastern quadrant (SEQ) and a significant lack of power in the northeastern quadrant (NEQ) in the Planck data. Asymmetries between the SEQ and the other three quadrants (southwestern quadrant (SWQ), northwestern quadrant (NWQ), and NEQ) are each in disagreement with an isotropic sky at a 95\% confidence level.  Furthermore, by rotating the Planck temperature sky map with respect to z direction, we showed the rotation angle where the TPCF of the SEQ has its maximal power. } 
{ Our analysis of the Planck foreground-cleaned maps shows that there is an excess of power in the TPCF in the SEQ and a significant  lack of power in the NEQ when compared with simulations.  This asymmetry is anomalous when considering the $\Lambda$CDM framework . }


\maketitle

\section{Introduction}

The standard cosmological model, also known as $\Lambda$CDM model, assumes homogeneous and isotropic expanding universe. At very early times, the universe, driven by a scalar field named inflaton, expanded rapidly. During this epoch, known as cosmological inflation, quantum fluctuations of the inflaton field were amplified to macroscopic scales, leading to the formation of primordial density perturbations, which then seeded the growth of cosmic structures. In this scenario, the perturbations are distributed as a homogeneous and isotropic Gaussian random field. At the epoch of decoupling between matter and radiation ($z \sim 1000$), primordial perturbations left an imprint in the CMB temperature as angular anisotropies. Deviations from statistical isotropy not related to foreground residuals or systematic effects can then point to interesting physical effects taking place in the early universe, and deserve further analysis.  

The first report of anomalies in the CMB temperature angular distribution dates back to the Cosmic Background Explorer (COBE) data release \citep{1992smoot} when the amplitude of the quadrupole value was found to be smaller than what was expected according to the standard cosmological model. Later, the low quadrupole amplitude was confirmed using the Wilkinson Microwave Anisotropy Probe (WMAP) data \citep{2003bennett.1, 2007hinshaw,2009hinshaw,2011jarosik}.  This result attracted much interest and resulted in further detailed statistical analysis of the CMB temperature angular distribution.  Other unexpected features then emerged from the data, with examples including an alignment of the low order multipoles \citep{2004bielewicz,2004schwarz,2004copi,2004deoliveiracosta,2005bielewicz,2005land,2006copi,2006abramo,2010frommert,2010gruppuso}, the cold spot,  which corresponds to a region with a significant  temperature decrement \citep{2004vielva,2005cruz,2007cruz,2010vielva.2}. More recently, the lack of large angle temperature correlations found previously on WMAP data were confirmed in Planck data by \citet{2013copi}.The anomaly more related to this paper are the north-south asymmetry \citep{2004eriksen,2004hansen,2004eriksen.2,2004hansen.2,2005donoghue,2009hoftuft,2010paci,2010pietrobon,2010vielva} and the quadrant asymmetry \citep{2012santos}. 

A follower of COBE and WMAP, the Planck satellite is a mission designed to measure temperature and polarization anisotropies of the CMB, with unprecedented sensitivity and angular resolution. Planck data can be used to search for possible systematic errors in WMAP results. Some of the anomalies in the angular distribution of CMB temperature found in the WMAP data are already shown to be present in the Planck data \citep{2013planck} in opposition to the \citet{2013bennett} claim that the anomalies are associated with systematics present in the data processing of WMAP first, third, fifth, and seventh year release. In addition, \citet{2011bennett} found no statistical significance for the reported anomalies in the CMB temperature fluctuations  found in WMAP data.  It may then be interesting to repeat the statistical isotropy analysis done for WMAP, and previously for COBE, on Planck data.

The purpose of our study is to investigate the presence of quadrant asymmetry, which was detected in WMAP data by  \citet{2012santos}, in the Planck data using the two-point angular correlation function (TPCF). This result showed the existence of an excess of power, at angular distances between pixels above 90 degrees, in the southeastern quadrant (SEQ).  In Section 2, we present our method, which involves the generation of numerical simulations of the CMB sky to be compared with real data. In Section 3, we discuss the results of our investigation. Finally, in Section 4, we draw some conclusions.

\section{Method}

The dataset used for our analysis is the main Planck CMB temperature foreground cleaned map SMICA together with a confidence mask, i.e, the region outside which the foreground removal procedure is considered statistically robust. For comparison, we later used also the NILC and SEVEM maps. For a detailed description of the maps and the confidence masks see \cite{2013planckc}. We obtain our main results by calculating the TPCF for four quadrants in the foreground-cleaned CMB temperature map.  The TPCF is estimated as the average product between the temperature of all pairs of  pixels  in the masked map separated by an angular distance $\gamma$. It is then defined as:

\begin{equation}
c(\gamma)\equiv \langle T({\bf n_p}) T({\bf n_q})\rangle.
\end{equation}

The temperature fluctuations of the pixels $p$ and $q$ are written as $T({\bf n_p})$ and $T({\bf n_q})$, respectively.

The pixels $p$ and $q$ are defined by the coordinates ($\theta_p$, $\phi_p$) and ($\theta_q$, $\phi_q$), where $0^\circ\leq \phi \leq 360^\circ$ and $-90^\circ \leq \theta \leq 90^\circ$. The angular distance between two generic pixels is then:

\begin{equation}
\cos\gamma = \cos\theta_p \cos\theta_q + \sin\theta_p \sin\theta_q \cos(\phi_p-\phi_q).
\end{equation}

The TPCF derived from real data for each quadrant in the sky is then compared to the results for the simulated sky maps. In Galactic coordinates, we started our analysis by selecting the quadrants  delimited by the Galactic plane and the plane perpendicular to the Galactic equator. We progressively rotate the map with respect to the z  axis (south to north Galactic pole direction) from 0  to 5, 10, 30, and 60  degrees to determine possible changes in the TPCF power.  Finally, we compare the results obtained with Planck data to the results previously obtained using WMAP data. We summarize our procedure in a few steps as described below:

% started our analysis by selecting the quadrants  delimited by the Galactic plane(xy-plane) and the xz-plane.
\begin{itemize}
\item We generated 1000 Monte Carlo (MC) simulations, with $N_{side}=256$ (pixel size$= 14'$), using the HEALPix (Hierarchical Equal Area and Isolatitude Pixelization) package (Synfast) \citep{2005gorski} with the best-fit $\Lambda$CDM model power spectrum from Planck \citep{2013planckb} as an input.

\item We divided the Planck  foreground-cleaned maps (SMICA, NILC, and SEVEM, taken from\cite{2013planckb}) and the simulated maps, expressed in Galactic coordinates, into quadrants:southeastern (SEQ), southwestern (SWQ), northeastern (NEQ) and northwestern (NWQ);

\item Even though the foreground-cleaned CMB maps reconstruct the CMB field over most of the sky, residual contamination in a stripe close to the Galactic plane is still present \citep{2013planckc}. Therefore, to ensure that our results are cosmological and not due to foreground contamination, we minimize the impact of the residual Galactic and point sources foregrounds by masking the contaminated pixels. We used the main confidence mask provided by the Planck team, which cuts 16.3\% of the sky for $N_{side}=64$. To see how sky cuts influence the TPCF read, \citet {2013gruppuso}.;

\item We removed the residual monopole and dipole from the maps using HEALPix packages (Anafast and Synfast).

\item We then calculate the TPCF for each quadrant of every map: since fluctuations at small angular scales are not relevant for our purposes, we degraded each simulated map as well as the used Planck maps and mask to $N_{side}=64$ in order to speed up the analysis. In the last case, for the $N_{side}=64$, a mask value equal to or smaller than $0.5$ means that the pixel is rejected. 

%Furthermore,  \textbf{We did not generated any noise for the simulations since the noise in the SMICA map is at least 2 orders of magnitude smaller than the signal for angular scales bigger than 1 degree \citep{2013planckb}};

\item Finally, for each TPCF curve, we calculated a rms-like quantity, $\sigma$, defined as \citep{2006bernui}:
\begin{equation}
\label{sigma}
	\sigma = \sqrt{\frac{1}{N_{bins}}\sum_{i=1}^{N_{bins}}f_i^2},
\end{equation}
\noindent where the $f_i$ corresponds to the TPCF  for each bin $i$. 

\end{itemize}

%We present the results for the Planck first release CMB temperature distribution dataset . We calculated the TPCF for the main ILC map smica \citep{2013planckb}. We divided the map in 4 quadrants considering the galaxy direction as the zero degree rotation. As a consistency test, we repeated the procedure for the nilc and sevem maps \citep{2013planckb}.  To avoid residual Galactic foreground we applied the mask provided by Planck team to the maps.

%We compare the results obtained with the data to monte carlo (MC) simulations generated according to the best fit $\Lambda$CDM model power spectrum using HEALPix (hierarchical equal area and isolatitude pixelization) package (synfast) \citep{2005gorski}.  We produced 1000 simulated maps with $N_{side}=256$.

The number of pixels in each quadrant after the cut does not differ substantially, being 10006, 10600,10578, 9955 for the SEQ, SWQ, NEQ, and NWQ, respectively. We used a number of bins $N_{bins}=90$ for our calculations, resulting in an angular distance between different samples of 2 degrees.

To quantify the statistical significance level of the asymmetry between quadrants in the Planck  foreground-cleaned maps, we compare the observed $\sigma$ values from Planck data with the mean $\sigma$ values measured from the 1000 $\Lambda$CDM sky simulations for each  quadrant . Finally, we search for the number of MC simulations with $\sigma$ values higher or smaller than those found on the Planck temperature map for each quadrant. In this way, we are able to see if there is any excess or lack of power and if the temperature fluctuation of the quadrants are consistent with the $\Lambda$CDM model.

%Finally, by calculating the TPCF and $\sigma$ values standard deviation for the MC simulations, we are able to see if Planck temperature fluctuation is consistent with the $\Lambda$CDM model.

%The comparison of the $\sigma$ values extracted from real data with the ones from simulations allows us to asses the statistical significance of the results. 

%\subsection{Two-point angular correlation function} \label{function}

%The TPCF is defined as
 
%\begin{equation}
%c(\gamma)\equiv \langle T({\bf n_p}) T({\bf n_q})\rangle.
%\end{equation}

%\noindent $T({\bf n_p})$ and $T({\bf n_q})$ corresponds to the temperature fluctuations of the pixels $p$ and $q$, respectively. We call the angular distance between two pixels  $\gamma$.

%The pixels $p$ and $q$ are defined by the coordinates ($\theta_p$, $\phi_p$) and ($\theta_q$, $\phi_q$), where $0^\circ\leq \phi \leq 360^\circ$ and $-90^\circ \leq \theta \leq 90^\circ$.   $\gamma$ is then defined as

%\begin{equation}
%\cos\gamma = \cos\theta_p \cos\theta_q + \sin\theta_p \sin\theta_q \cos(\phi_p-\phi_q).
%\end{equation}

%Defining the quantity $\sigma$ (rms-like), we can compare the TPCF computed both for the MC simulations and for Planck  data \citep{2006bernui}:

%\begin{equation}
%\label{sigma}
%	\sigma = \sqrt{\frac{1}{N_{bins}}\sum_{i=1}^{N_{bins}}f_i^2}. 
%\end{equation}

%The $f_i$ corresponds to the TPCF  for each bin $i$. We used the number of bins $N_{bins}=90$ to quantify our results.

\section{Results and discussion}

\begin{figure}
 \includegraphics[scale=0.5]{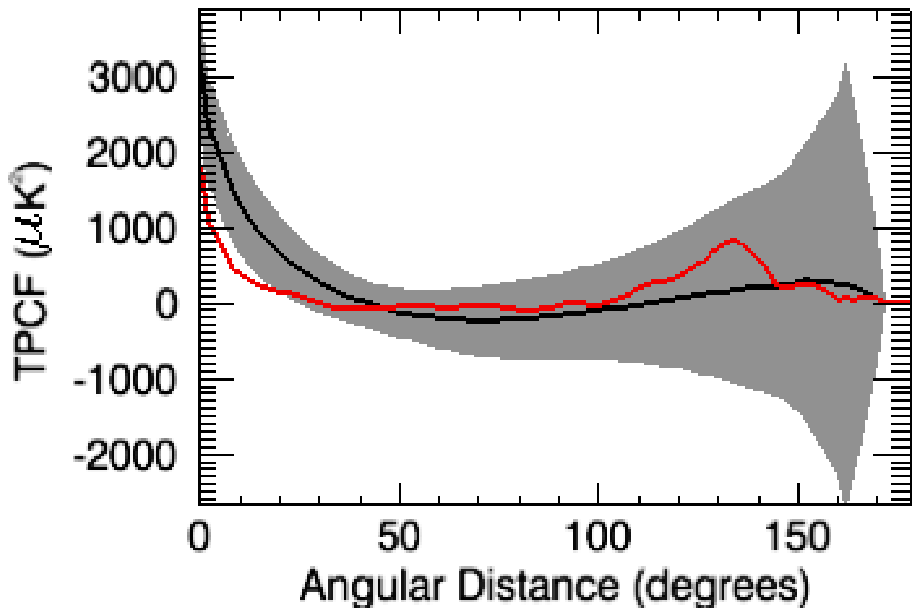}
 \includegraphics[scale=0.5]{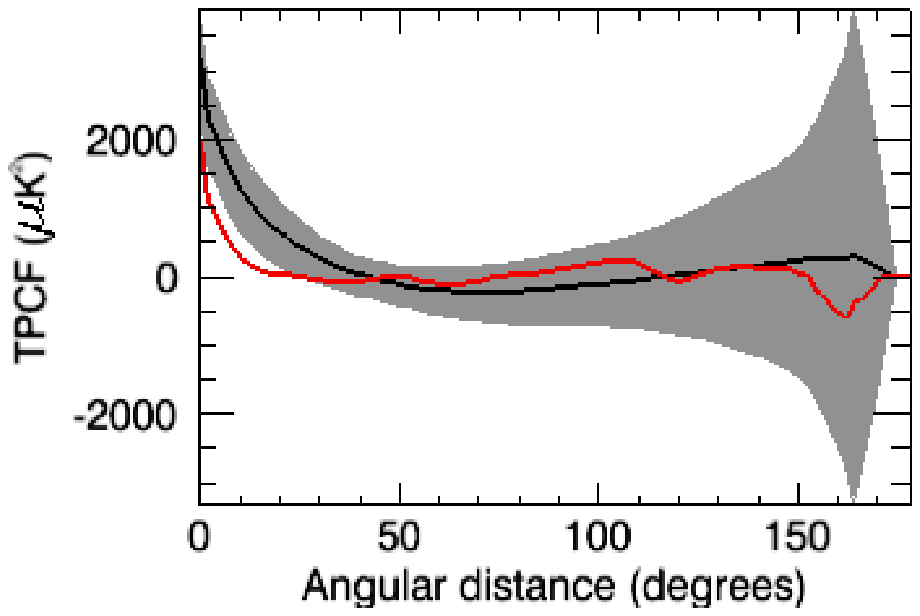}
 \includegraphics[scale=0.5]{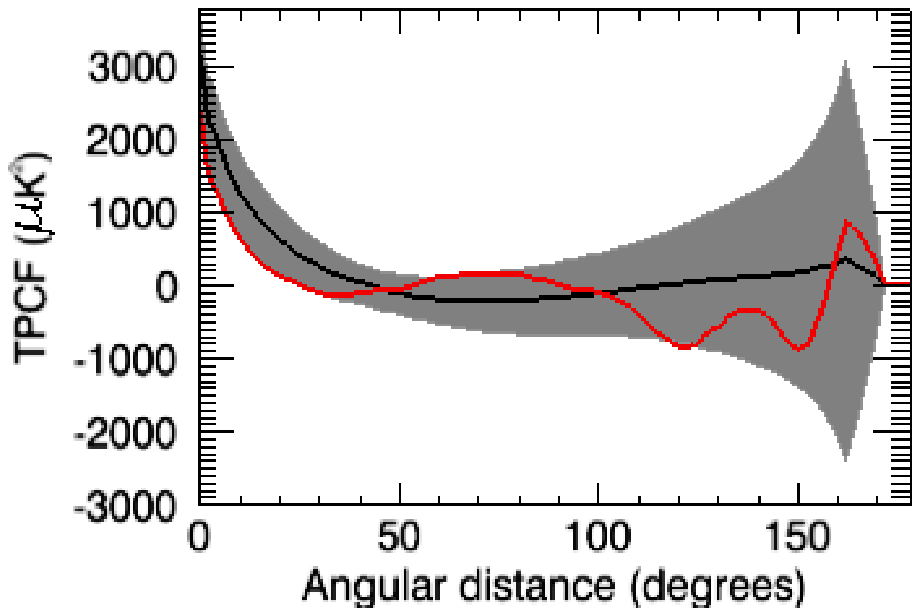}
 \includegraphics[scale=0.5]{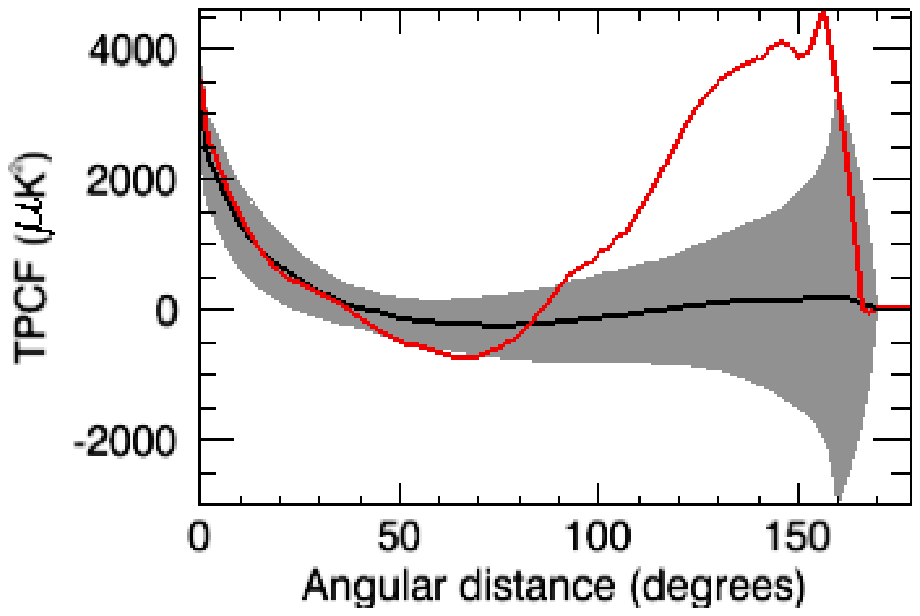}
\caption{TPCF curves computed for the main Planck  temperature  foreground-cleaned map (SMICA) (red curve) also using the mask provided by Planck team.  We smoothed the curves using the smooth function from Interactive Data Language (IDL) for illustration purposes only (in the calculations we use the original calculated values for the TPCF). From top to bottom, NWQ, NEQ, SWQ, and SEQ appear as solid red lines. The shadow part depicts the standard deviation intervals  (68\% C.L)  for 1000 simulated maps produced with the $\Lambda$CDM spectrum. The black curve is the mean TPCF considering the MC simulated maps.}
\label{TPCF-NWQ-NEQ}
\end{figure}

Our analysis confirmed an excess of power in the SEQ Planck temperature data as shown in Figure \ref{TPCF-NWQ-NEQ}. The TPCF for this quadrant is outside the standard deviation for the TPCF found in the 1000 MC simulations for the best fit of the standard cosmological model. For the other quadrants, the curves corresponding to the data lay inside the MC standard deviation (Figure \ref{TPCF-NWQ-NEQ}) %and \ref{TPCF-SWQ-SEQ}). 

\begin{figure}
\includegraphics[scale=0.6]{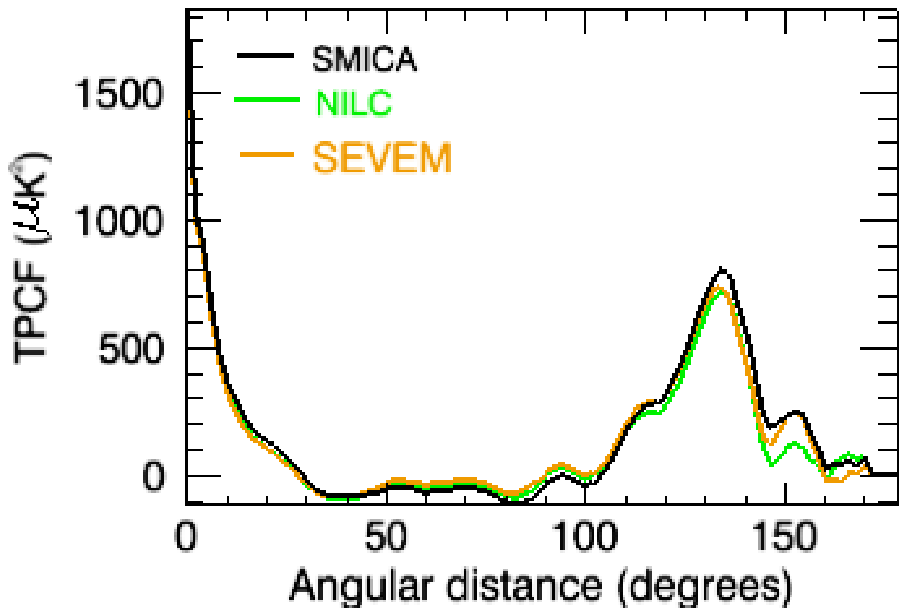}
\includegraphics[scale=0.6]{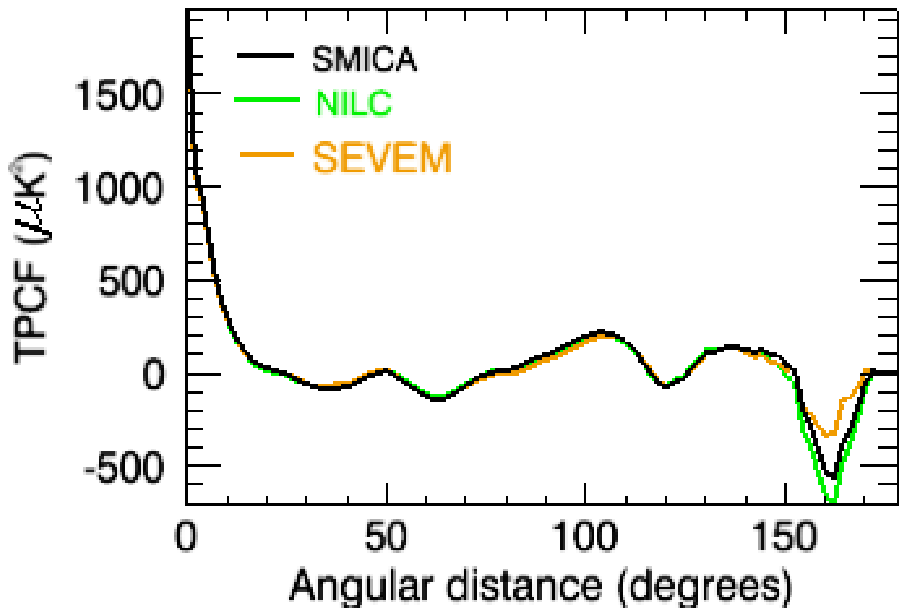}
\includegraphics[scale=0.6]{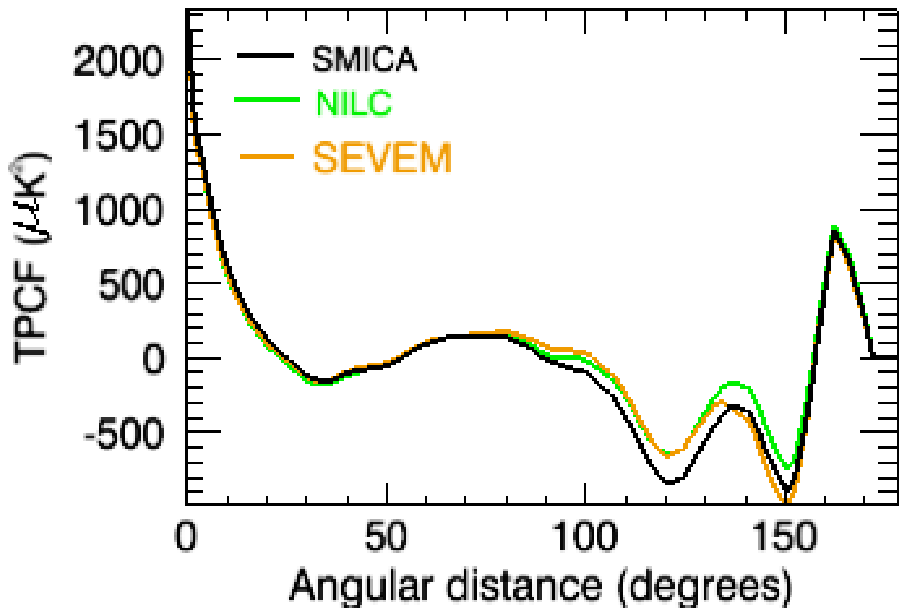}
\includegraphics[scale=0.6]{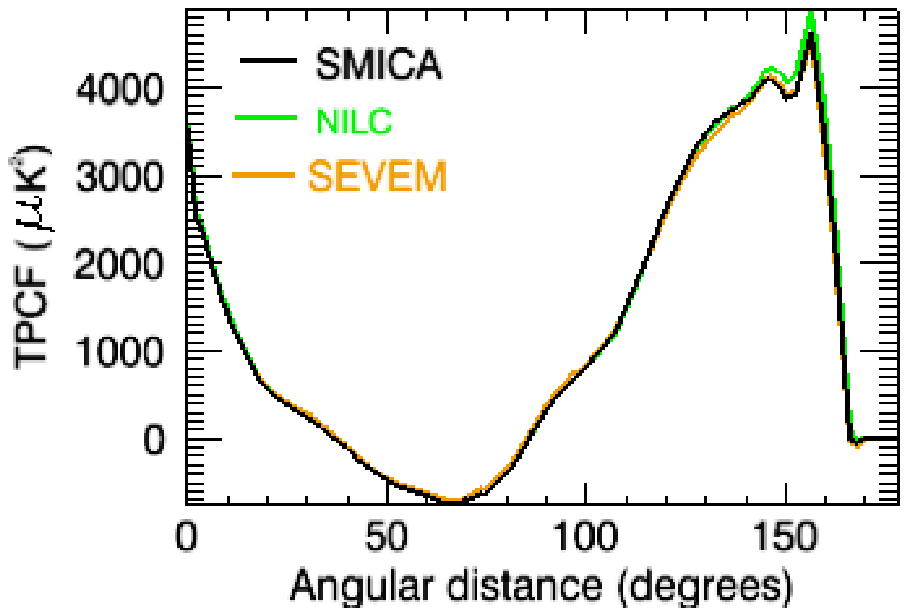}
\caption{Comparison between the Planck  foreground-cleaned maps (SMICA, NILC and SEVEM). From  top to bottom, the curves refer to the NWQ, NEQ, SWQ, and SEQ}
\label{TPCF-NWQ-NEQ-nilc}
\end{figure}

\begin{figure}
\includegraphics[scale=0.6]{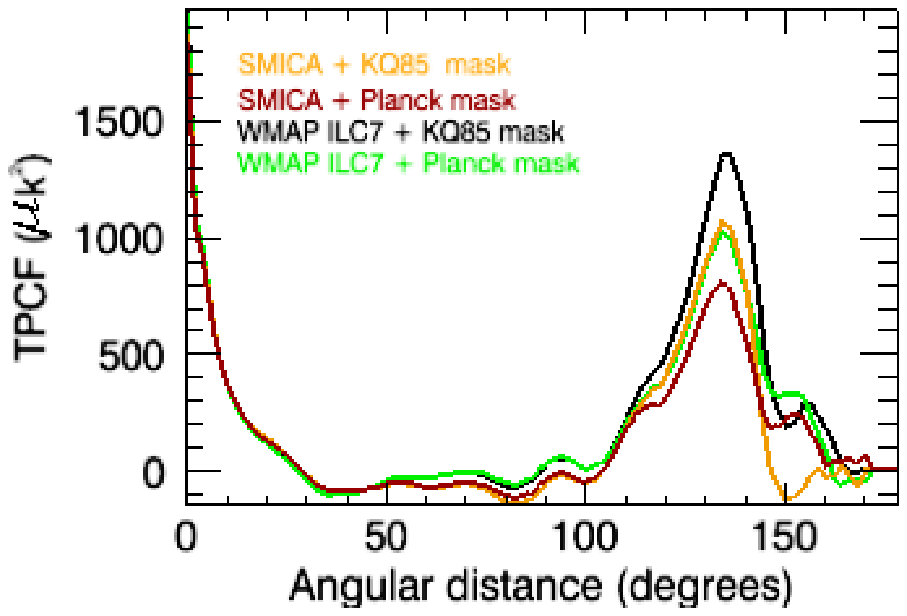}
\includegraphics[scale=0.6]{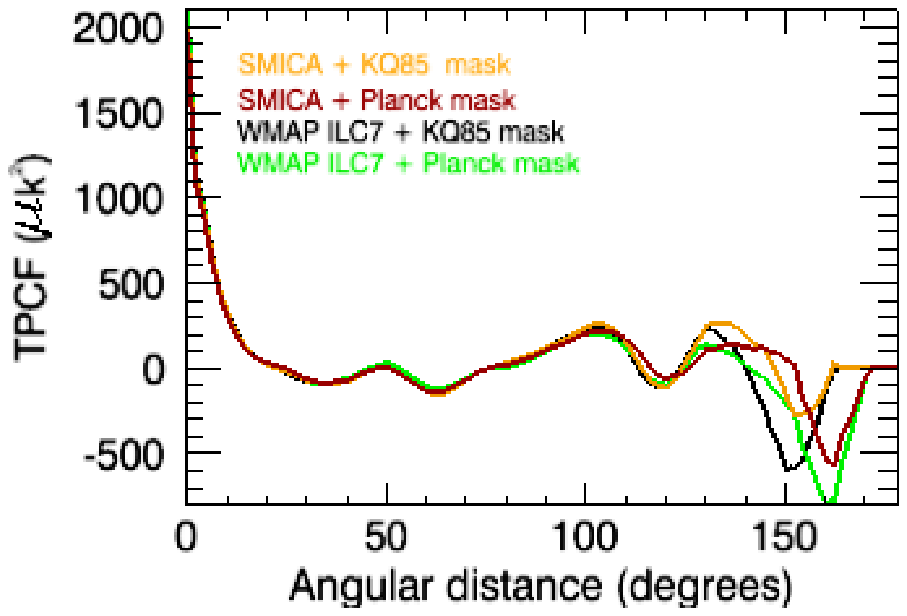}
\includegraphics[scale=0.6]{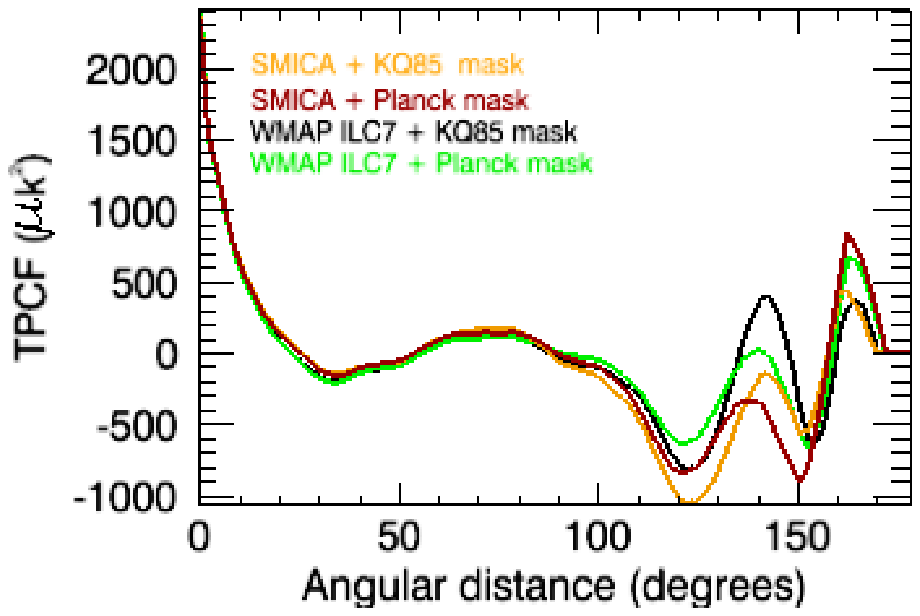}
\includegraphics[scale=0.6]{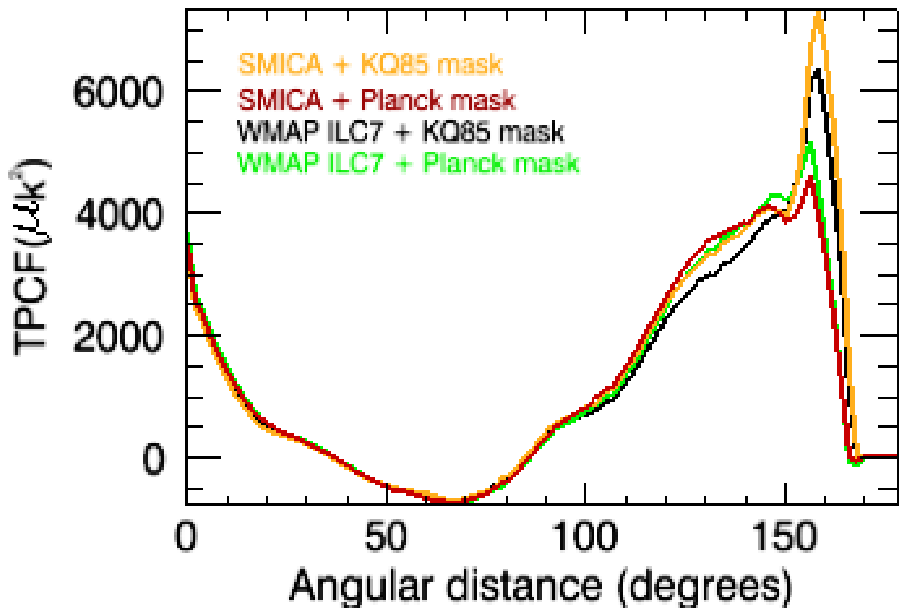}
\caption{Comparison between NWQ, NEQ, SWQ and SEQ (from top to bottom) for the TPCF using the WMAP KQ85 mask and the Planck mask for both  maps:  ILC WMAP7 and  foreground-cleaned SMICA.} 
 \label{TPCF-mask-masknova1}
\end{figure}

For the sake of comparison, we also calculated the TPCF for the NILC and SEVEM maps, finding results consistent with the result using the main  foreground-cleaned Planck map SMICA for each quadrant (see Figure \ref{TPCF-NWQ-NEQ-nilc}).  As a summary, we show the TPCF for WMAP 7-year ILC map and SMICA using both Planck and WMAP masks for both maps in Figure \ref{TPCF-mask-masknova1}.  We see that the results agree among each other as expected.

%\begin{figure}
%\resizebox{\hsize}{!} {\includegraphics[scale=0.3]{./figuras/QSE_planck.eps}}
%\resizebox{\hsize}{!} {\includegraphics[scale=0.3]{./figuras/QSD_planck.eps}}
%\resizebox{\hsize}{!} {\includegraphics[scale=0.3]{./figuras/QIE_planck.eps}}
%\resizebox{\hsize}{!} {\includegraphics[scale=0.3]{./figuras/QID_planck.eps}}
%\caption{TPCF curves computed for the main Planck  temperature  foreground-cleaned map (SMICA) (red curve) using  also the mask provided by Planck team.  We smoothed the curves using the smooth function from Interactive Data Language (IDL) for illustration purposes only (in the calculations we use the original calculated values for the TPCF). The NWQ (top) and the NEQ (bottom) appear as solid red lines. The shadow part depicts the standard deviation intervals  (68\% C.L)  for 1000 simulated maps produced with the $\Lambda$CDM spectrum. The black curve is the mean TPCF considering the MC simulated maps.}
%\label{TPCF-NWQ-NEQ}
%\end{figure} 

%\begin{figure}
%\resizebox{\hsize}{!} {\includegraphics{./figuras/QIE_planck.eps}}
%\resizebox{\hsize}{!} {\includegraphics{./figuras/QID_planck.eps}}
%\caption{Same as Figure \ref{TPCF-NWQ-NEQ}, but now with the SWQ at the top and the SEQ at the bottom.}
%\label{TPCF-SWQ-SEQ}
%\end{figure} 

%\begin{figure}
%\resizebox{\hsize}{!} {\includegraphics{./figuras/QSE_planck_nilc_sevem.eps}}
%\resizebox{\hsize}{!} {\includegraphics{./figuras/QSD_planck_nilc_sevem.eps}}
%\caption{Comparison between the Planck  foreground-cleaned maps (SMICA, NILC and SEVEM). From  top to bottom, the curves refer to the NWQ and the NEQ, respectively.}
%\label{TPCF-NWQ-NEQ-nilc}
%\end{figure} 

%\begin{figure}
%\resizebox{\hsize}{!} {\includegraphics{./figuras/QIE_planck_nilc_sevem.eps}}
%\resizebox{\hsize}{!} {\includegraphics{./figuras/QID_planck_nilc_sevem.eps}}
%\caption{Same as Figure \ref{TPCF-NWQ-NEQ-nilc}. From top to bottom, the curves refer to the SWQ and the SEQ, respectively.}
%\label{TPCF-SWQ-SEQ-nilc}
%\end{figure} 

%\begin{figure}
%\resizebox{\hsize}{!}{\includegraphics{./figuras/QSE_masks.eps}}
%\resizebox{\hsize}{!}{\includegraphics{./figuras/QSD_masks.eps}}
%\caption{Comparison between NWQ (top) and NEQ (bottom)  for the TPCF using the WMAP KQ85 mask and the Planck mask for both  maps:  ILC WMAP7 and  foreground-cleaned SMICA.} 
 %\label{TPCF-mask-masknova1}
%\end{figure}

%\begin{figure}
%\resizebox{\hsize}{!}{\includegraphics{./figuras/QIE_masks.eps}}
%\resizebox{\hsize}{!}{\includegraphics{./figuras/QID_masks.eps}}
%\caption{Same as Figure  \ref{TPCF-mask-masknova1} for the SWQ (top) and the SEQ (bottom). } \label{TPCF-mask-masknova2}
%\end{figure}

\begin{figure}
\includegraphics[scale=0.6]{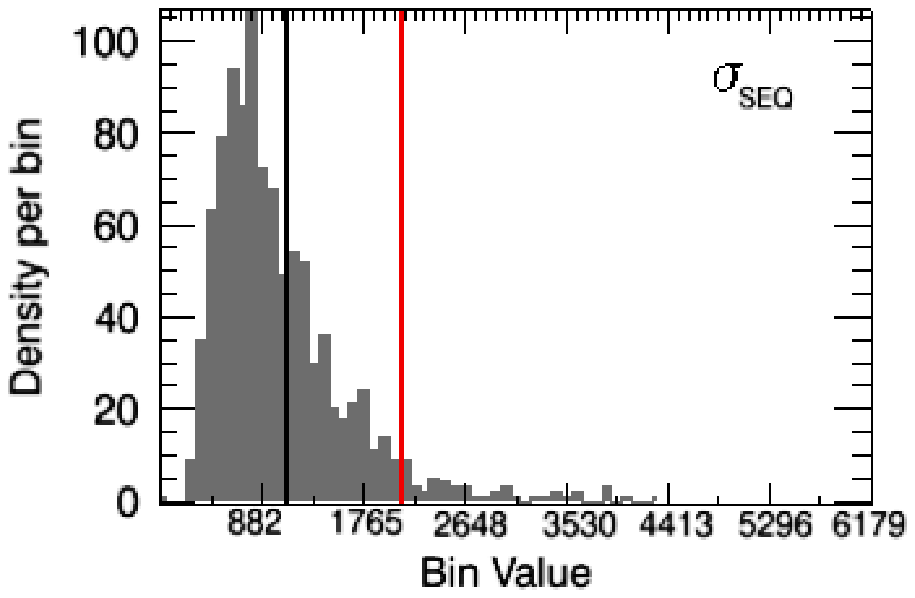}
\includegraphics[scale=0.6]{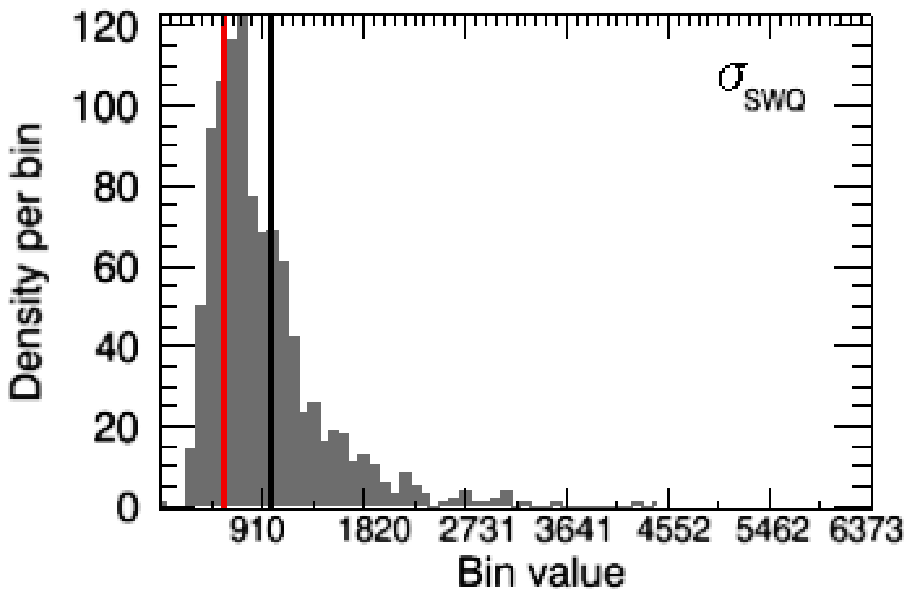}\\
\includegraphics[scale=0.6]{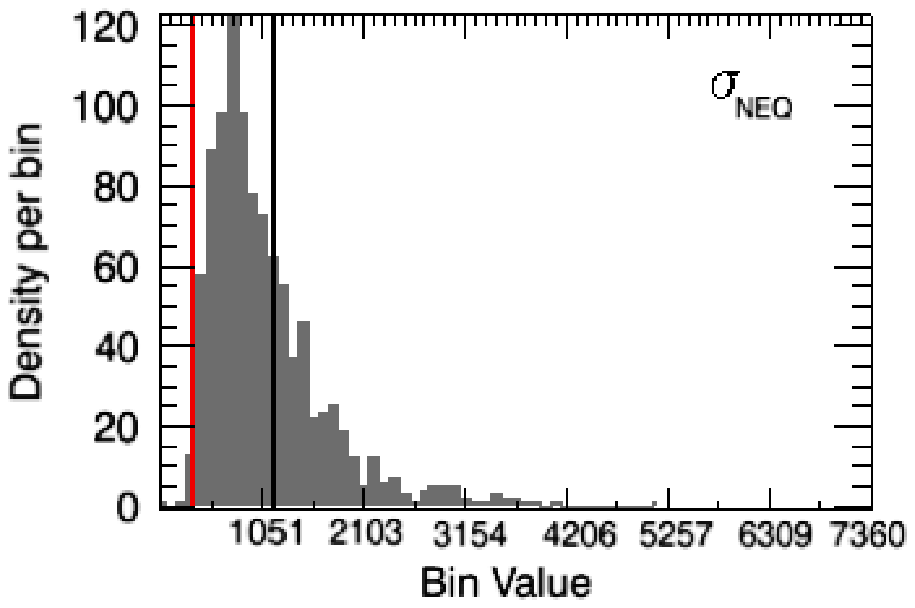}
\includegraphics[scale=0.6]{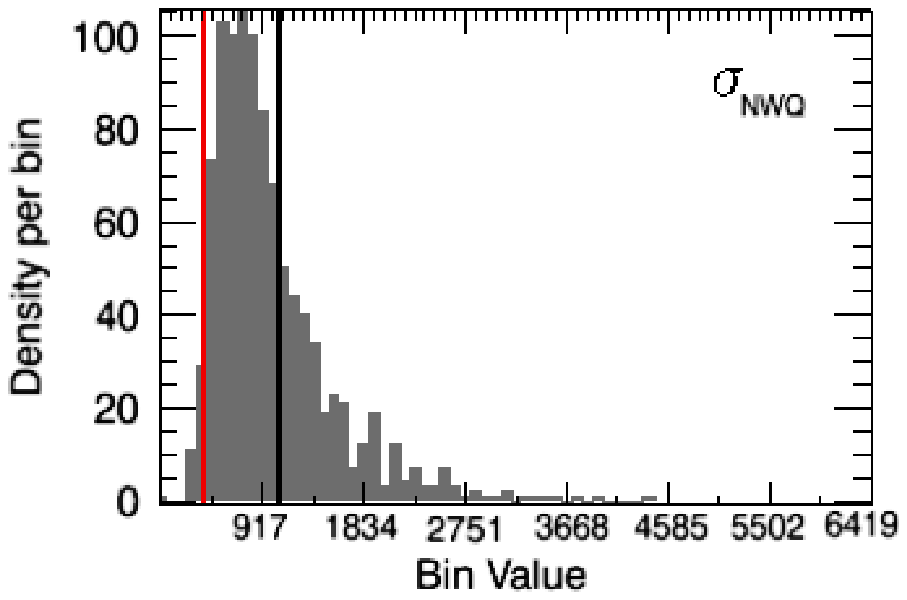}
\caption{Distribution of the $\sigma$ values for each quadrant of the MC simulations. The black vertical line indicates where the mean value is located. The red line stands for the value found in SMICA + Planck mask.}
\label{hist_quadrant}
\end{figure}

We quantify the TPCF curves using the definition of Equation \ref{sigma}. We found that, for the SEQ, 19.3\% of the MC simulations present at least one quadrant with $\sigma$ value equal to or higher than  that value found in the SMICA Planck map. In the same manner, we found a lack of large-angle temperature correlation in the other quadrants considering the $\Lambda$CDM model. Comparing the SWQ $\sigma$ values from the data with those obtained from simulations, we found 32.9\% of the simulated maps with a quadrant having $\sigma$ values equal to or smaller than that of the SMICA SWQ. In the northern quadrant, the lack of power in the data becomes even more evident, being $\sigma$ smaller or equal to the value found in the data in 2.4\% and 0.5\% of the MC simulations for the NWQ and NEQ, respectively (see Table \ref{tbl-prob-quadrants} and Figure \ref{hist_quadrant} for a summary of the results).   The fact that the two point correlation function for the northern hemisphere is featureless relative to the southern hemisphere is in agreement with the results found in WMAP data \citep{2004eriksen} and confirmed by analyzing Planck temperature maps \citep{2013planck}.

\begin{figure}
\includegraphics[scale=0.6]{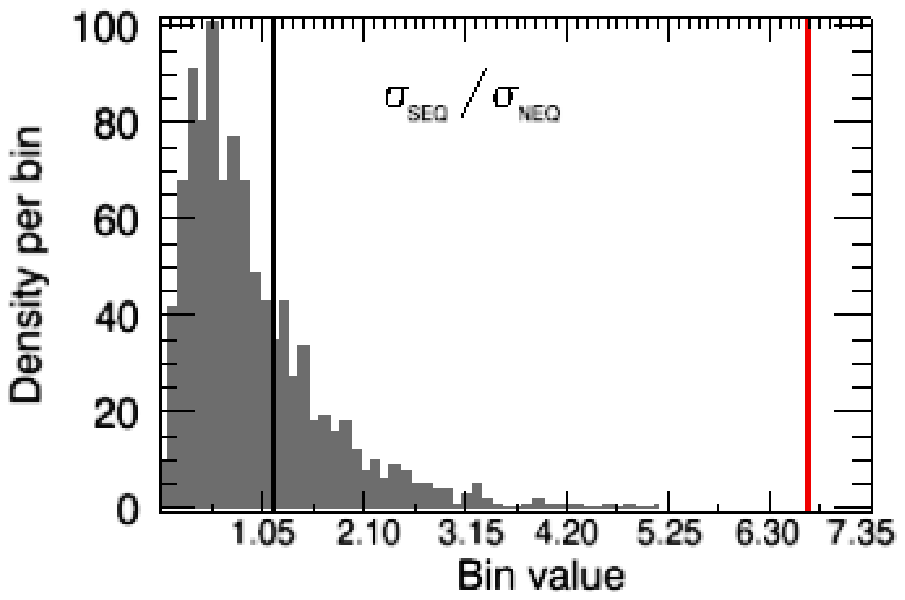}
\includegraphics[scale=0.6]{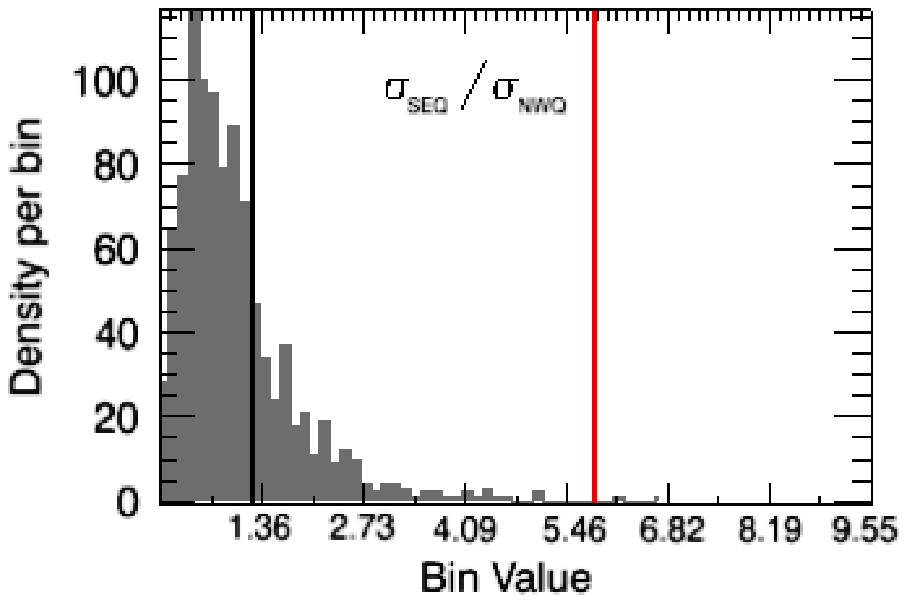}
\includegraphics[scale=0.6]{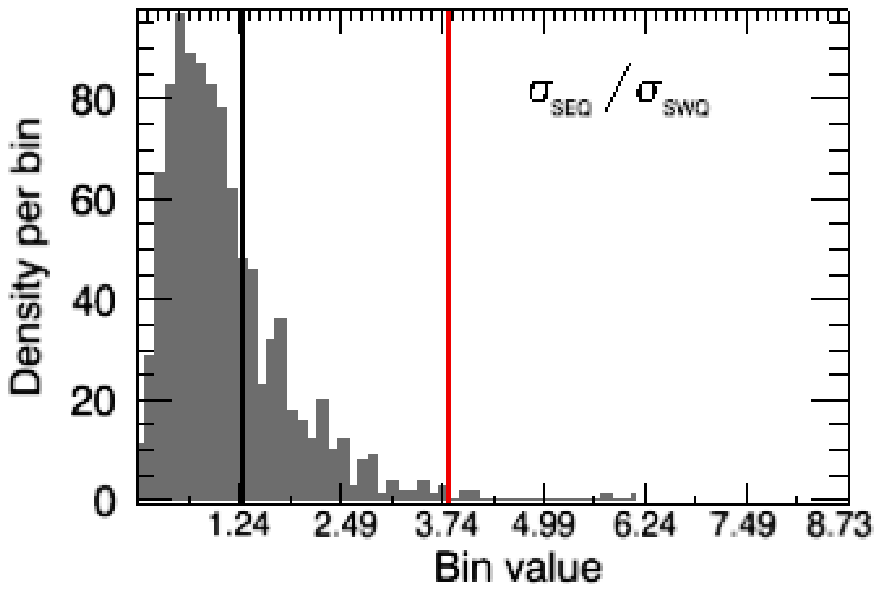}
\caption{Distribution of the $\sigma$ ratio between pair of quadrants. The black vertical line indicates where the mean value lies in the Monte Carlo simulations. The red line stands for the value found in SMICA + Planck mask.}
\label{hist}
\end{figure}

\begin{figure}
\includegraphics[scale=0.6]{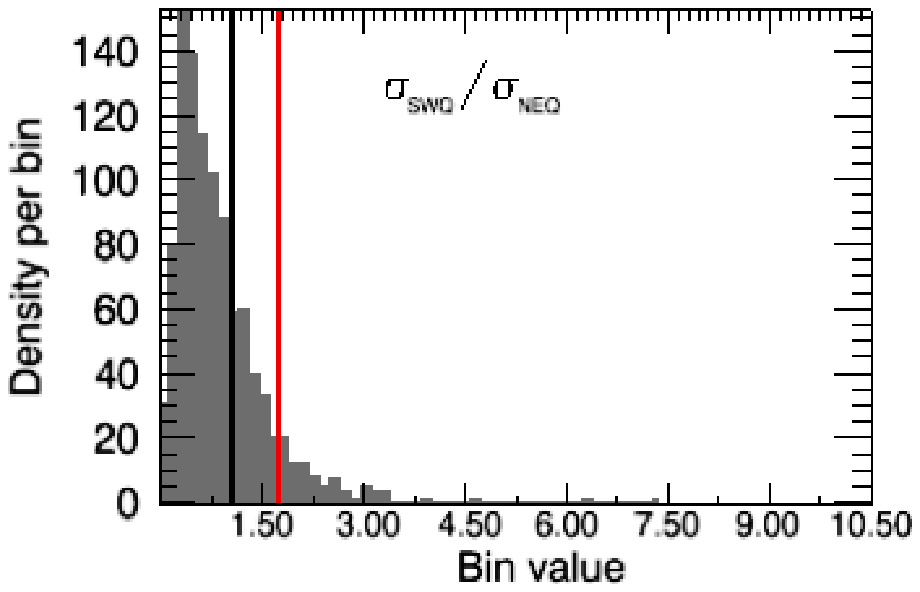}
\includegraphics[scale=0.6]{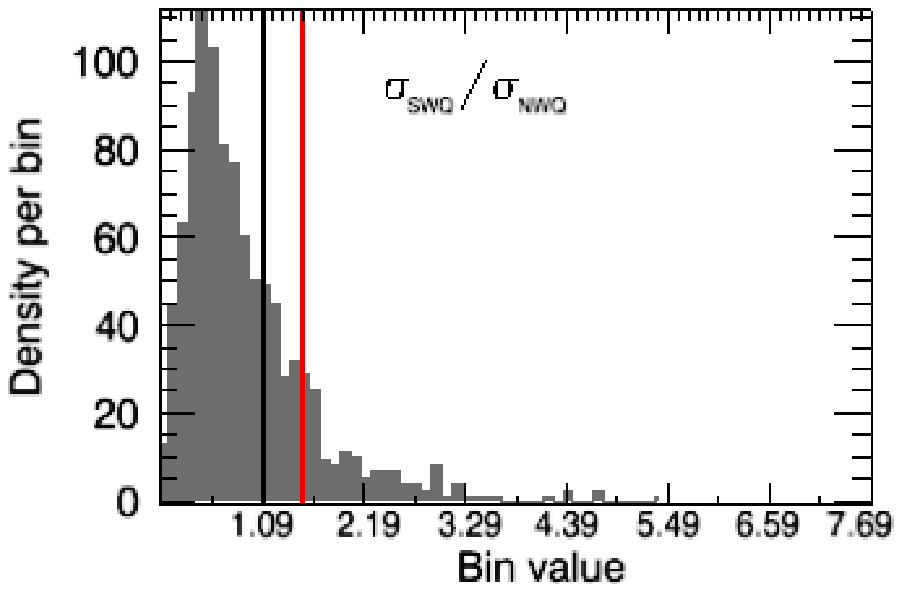}
\includegraphics[scale=0.6]{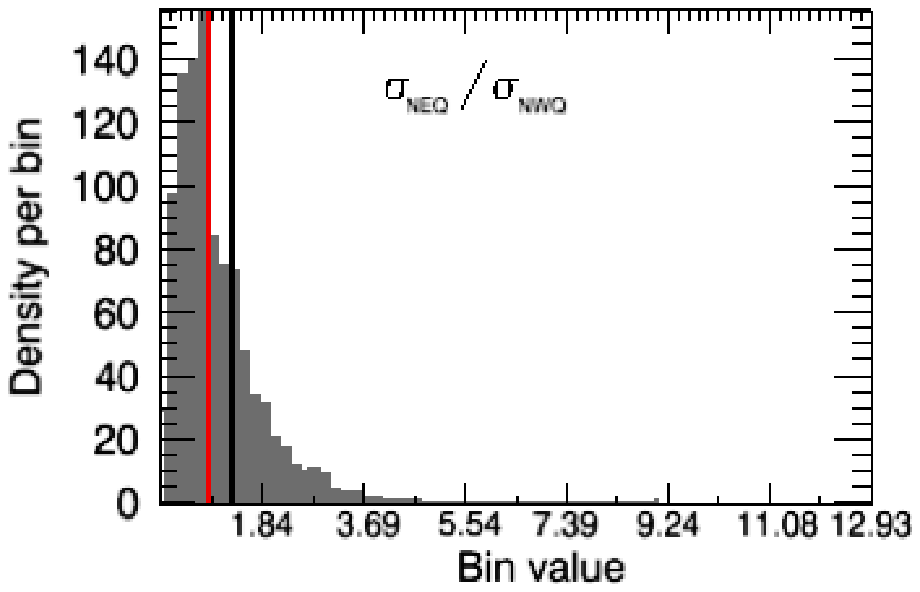}
\caption{Same as Figure \ref{hist} for different pairs of quadrants.}
\label{hist1}
\end{figure}

Furthermore, it is important to compare the excess power found in the SEQ with the lack of power in the other quadrants, specially in the NEQ.  Since the universe is expected to be isotropic, according to the standard cosmological model, we would not expect to find  large deviations from 1 for $\sigma_i/\sigma_j$  between \textit{i} and \textit{j} quadrants (\textit{i}, \textit{j}=\{SEQ, SWQ, NWQ, NEQ\}). This result is confirmed by the mean found for the ratios between the  $\sigma$ values in each quadrant in the MC simulations, as we can see in Table \ref{tbl-sigma} and Figures \ref{hist} and \ref{hist1} , where we emphasize the SEQ with respect to the others since this quadrant has an excess of power in the real data. The mean values that slightly diverge from one\footnote{It is important to point out that we used a different method from \citet{2012santos}, found more appropriate,  to calculate the mean values of the $\sigma$ ratios.}, for example the  $\sigma_{SEQ}/\sigma_{NWQ}$, are due to the different number of pixels in each quadrant as a result of the use of the mask.  In Table \ref{tbl-sigma}, we show the mean values of the rate between the quantity $\sigma$ for each of the mentioned quadrants and its correspondent 68\%, 95\%, and 99.7\% C.L. considering 1000 MC simulations.

%\begin{table}
%\begin{center}
%\caption{ Mean of the sigma ratios using the simulated CMB maps considering the $\Lambda$CDM model. The errors are 68\% confidence level. \label{tbl-sigma} }
%\begin{tabular}{ccc}
%\hline\hline
%$\sigma_{SEQ}/\sigma_{NWQ}$ &$\sigma_{SEQ}/\sigma_{NEQ}$ & $\sigma_{SEQ}/\sigma_{SWQ}$\\
%\hline
%$ 1.02^ {+0.62}_ {-0.39}$ &$ 0.94^ {+0.59}_ {-0.34}$ &$ 1.1^ {+0.62}_ {-0.42}$\\ 
%\hline\hline
%$\sigma_{SWQ}/\sigma_{NEQ}$ &$\sigma_{SWQ}/\sigma_{NWQ}$ &$\sigma_{NEQ}/\sigma_{NWQ}$\\
%\hline
%$0.85^{+ 0.51}_{-0.27}$ &$0.92^{+ 0.55}_{-0.32}$ &$1.1^{+0.61}_{-0.41}$\\
%\hline\hline 
   %\end{tabular}
%\end{center}
%\end{table}

%\begin{table}
%\begin{center}
%\caption{Mean, 68\% and 95\% C.L.  values of the sigma ratios for different pair of quadrants using the simulated CMB maps considering the $\Lambda$CDM model. \label{tbl-sigma} }
%\begin{tabular}{cccccc}
%\hline\hline
 %            &95\% C.L. & 68\% C.L. & mean &68\% C.L. &95\% C.L.\\
 %\hline                                                           
%$\sigma_{SEQ}/\sigma_{NWQ}$  &0.42&0.82&1.83&3.06&4.88 \\
%\hline
%$\sigma_{SEQ}/\sigma_{NEQ}$ &0.39&0.76&1.14&2.48&3.82\\ 
%\hline
%$\sigma_{SEQ}/\sigma_{SWQ}$&0.44&0.86&1.29&2.90&4.52\\
%\hline
%$\sigma_{SWQ}/\sigma_{NEQ}$&0.35&0.69&1.02&3.14&5.25\\
%\hline
%$\sigma_{SWQ}/\sigma_{NWQ}$&0.38&0.74&1.10&2.53&3.97 \\
%\hline
%$\sigma_{NEQ}/\sigma_{NWQ}$&0.44&0.86&1.27&3.88&6.48 \\
%\hline\hline 

%\end{tabular}
%\end{center}
%\end{table}

\begin{table}
\begin{center}
\caption{Mean, 68.2\%, 95\%, and 99.7\% C.L.  values of the sigma ratios for different pair of quadrants using the simulated CMB maps considering the $\Lambda$CDM model. \label{tbl-sigma} }
\begin{tabular}{ccccc}
\hline\hline
             &mean &68\% C.L. &95\% C.L. & 99.7\% C.L.\\
 \hline                                                           
$\sigma_{SEQ}/\sigma_{NWQ}$  &1.83  &1.17 &2.58 &5.15 \\
\hline
$\sigma_{SEQ}/\sigma_{NEQ}$ &1.14    &1.13 &2.55 &4.45\\ 
\hline
$\sigma_{SEQ}/\sigma_{SWQ}$&1.29    &1.26 &2.60& 4.13\\
\hline
$\sigma_{SWQ}/\sigma_{NEQ}$&1.02    &1.01 &2.15 &4.05\\
\hline
$\sigma_{SWQ}/\sigma_{NWQ}$&1.10   &1.09 &2.40 & 4.58 \\
\hline
$\sigma_{NEQ}/\sigma_{NWQ}$&1.27   &1.22 &2.61 &4.43 \\
\hline\hline

\end{tabular}
\end{center}
\end{table}

The asymmetries found in the data are not present in 99\% of the MC simulated maps considering exactly the same configuration as for the data, as can be seen in Tables \ref{tbl-sigma} and \ref{tbl-prob-mask} (probability P1) and Figures \ref{hist} and \ref{hist1}.  We calculated the probability that the values for $\sigma_{SEQ}/\sigma_j$, being $\textit{j}=\{SWQ, NWQ, NEQ\})$, in the simulated maps have equal to or higher values than those found in the Planck data. We found that 1\%, 0.2\%, and less than 0.1\% of the simulations have the asymmetries presented in the SMICA map for $\sigma_{SEQ}/\sigma_{SWQ}$, $\sigma_{SEQ}/\sigma_{NWQ}$ and $\sigma_{SEQ}/\sigma_{NEQ}$, respectively.  As we can  also see in Table \ref{tbl-prob-mask}, the largest asymmetry occurs between the SEQ and the NEQ, the same found by \cite{2012santos} in WMAP data using the KQ85 mask. Taking into consideration that the occurrence of this asymmetry in the data ($\sigma_{SEQ}/\sigma_{NEQ}$) can happen between any quadrant in the simulated maps, regardless of the different number of pixels in each quadrant because of the use of the mask, we found a probability of 0.8\% for this asymmetry to happen (see Table \ref{tbl-prob-mask}, probability P3).  It is important to point out that \cite{2012santos} found no evidence for a connection between the SEQ excess of power and the so-called cold spot anomaly (see  \cite{2004vielva,2005cruz,2007cruz,2010vielva.2}).

%Even though smaller compared to the others, the asymmetry between the SEQ and the SWQ is still outside a 99 \% confidence level.  

On the other hand, as expected, $\sigma_i/\sigma_{NEQ}$, $\sigma_i/\sigma_{NWQ}$ and $\sigma_i/\sigma_{SWQ}$, in this case for (\textit{i}=\{SWQ, NWQ, NEQ\}) are nearly 1 and well inside the MC 99.7\% C.L. (see Table \ref{tbl_sigma_planck}). In  Table \ref{tbl_sigma_planck}, we can also see once again that the results are in agreement when WMAP KQ85 or Planck masks are used in the analysis.

%\begin{figure}
%\resizebox{\hsize}{!} {\includegraphics{./figuras/sigma_points_NEQ_3sig.eps}}
%\resizebox{\hsize}{!} {\includegraphics{./figuras/sigma_points_NWQ_3sig.eps}}
%\resizebox{\hsize}{!} {\includegraphics{./figuras/sigma_points_SWQ_3sig.eps}}
%\caption{The values of  $\sigma_i/\sigma_{NEQ}$, $\sigma_i/\sigma_{NWQ}$ and $\sigma_i/\sigma_{SWQ}$ for the data (red diamonds), the medium values considering the MC simulations (black asterisk) and the 99.7\% C.L error bars (gray solid line) for each quadrant fraction combination. The x-axis shows the \textit{i}=\{SEQ, SWQ, NWQ, NEQ\} quadrants.}
%\label{sigma_points}
%\end{figure} 

\begin{figure}
\resizebox{\hsize}{!}{\includegraphics{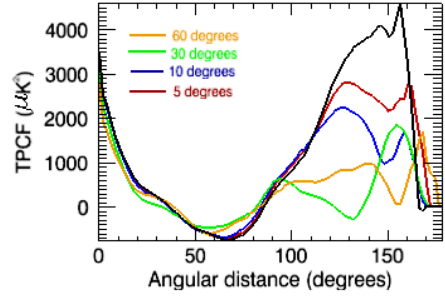}}
\resizebox{\hsize}{!}{\includegraphics{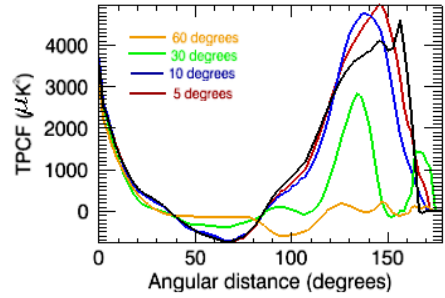}}
\caption{Top: Anti-clockwise rotation performed in the SEQ with respect to z direction for 60, 30, 10, and 5 degrees. Bottom: Clockwise rotation performed in SEQ with respect to z direction for 60, 30, 10, and 5 degrees. In both plots, the black curve corresponds to a 0-degree rotation.}
 \label{rotation} 
\end{figure}

Finally, we rotated the CMB temperature map with respect to the $z$ axis, both in clockwise and anti-clockwise directions, to quantify the angle where the TPCF from SEQ reaches its highest power. The result is shown in Figure \ref{rotation}. We see that the TPCF reaches its highest value for the 5-degree clockwise rotation. Comparing the $\sigma$ values for the SEQ 0-degree rotation and the SEQ 5-degree clockwise rotation, we found an increase of almost 3 \% in the last case. Gradually, the excess of power on the TPCF starts to decrease as we rotate the sky above 10 degrees, since we are mixing, in this way, the initial SEQ  and SWQ,  the last one showing no anomaly previously.  The rotations in the anti-clockwise direction show a more evident dissipation of the excess of power as the rotation angle increases. By rotating the quadrant we were not able to identify a possible source of the excess of power in the SEQ. 

% For rotations above 10 degrees, differences in the TPCF with respect to the curve without any rotation start to be more evident.  In both cases, we have a small shift  to smaller angular distances in the curves for a 20-degree rotation. In the case of a 30-degree rotation we have already a dissipation of the excess power in the TPCF, in both directions.

\section{Conclusion}

 We confirmed the existence of  asymmetries between the SEQ and the other three quadrants, more evidently the NEQ,  in the Planck CMB temperature map.The largest asymmetry, found between the SEQ and NEQ,  is in disagreement with the assumption of statistical isotropy of the standard cosmological model above 95\% C.L , considering that this asymmetry could happen between any quadrant in the simulated sky maps.  The largest asymmetries occurs between the SEQ and the NEQ and NWQ. This result suggests that the excess power found earlier in the southern hemisphere corresponds to an excess power in the SEQ, since the asymmetries between the SWQ and the NEQ or NWQ, even though slightly larger than the asymmetry between the NEQ and the NWQ, are well inside the 99.7\% C.L.  We did not find, in 1000 simulated maps, all three asymmetries occurring simultaneously using the exactly same configuration of the Planck map quadrants. Finally, by performing rotations in the temperature map of 5, 10, 30, and 60 degrees with respect to the $z$ direction in Galactic coordinates, we found that the  5-degree clockwise rotation corresponds to the configuration where the power in the SEQ reaches its maximum. 

Further investigation is needed to explain the source of these asymmetries in the data. To date, a proper explanation is still missing in the literature. 

The authors acknowledge the use of HEALPix packages, of the Legacy Archive for Microwave Background Data Analysis (LAMBDA) and  of the Planck Legacy Archive (PLA).  The development of Planck has been supported by: ESA; CNES, and CNRS/INSU-IN2P3-INP (France); ASI, CNR, and INAF (Italy); NASA and DoE (USA); STFC and UKSA (UK); CSIC, MICINN, and JA (Spain); Tekes, AoF, and CSC (Finland); DLR and MPG (Germany); CSA (Canada); DTU Space (Denmark); SER/SSO (Switzerland); RCN (Norway); SFI (Ireland); FCT/MCTES (Portugal); and the development of Planck has been supported by: ESA; CNES, and CNRS/INSU-IN2P3-INP (France); ASI, CNR, and INAF (Italy); NASA and DoE (USA); STFC and UKSA (UK); CSIC, MICINN, and JA (Spain); Tekes, AoF, and CSC (Finland); DLR and MPG (Germany); CSA (Canada); DTU Space (Denmark); SER/SSO (Switzerland); RCN (Norway); SFI (Ireland); FCT/MCTES (Portugal); and PRACE (EU). A description of the Planck Collaboration and a list of its members, including the technical or scientific activities in which they have been involved, can be found at http://www.sciops.esa.int/index.php?project=planck\&page\\
=Planck\_Collaboration. T. Villela acknowledges CNPq support through grant 308113/2010-1.

\clearpage
\onecolumn

\begin{table}
\begin{center}
\caption{Probabilities of finding in the MC simulations the $\sigma$ value higher than the value found in Planck data for the SEQ  and smaller than the values found for SWQ, NEQ, and NWQ in the Planck map.}
\label{tbl-prob-quadrants} 
\begin{tabular}{ccrcrcrcr}
\hline\hline
 Map &  $\sigma_{SEQ}$ & P1 \tablefootmark{a} & $\sigma_{SWQ}$ & P2\tablefootmark{b} & $\sigma_{NEQ}$ & P3\tablefootmark{c}& $\sigma_{NWQ}$ & P4\tablefootmark{d} \\
 \hline
 SMICA +  Planck mask &  2090.32 & 19.3\%    &543.54    &32.9\%    & 311.03  &  0.5\% & 359.67 & 2.4\% \\ 
 \hline
\end{tabular}
\end{center}
\tablefoottext{a}{Probability of finding $\sigma_{MC}>2090.32$ in any quadrant of the simulated maps.}
\tablefoottext{b}{Probability of finding $\sigma_{MC}<543.54$  in any quadrant of the simulated maps.}
\tablefoottext{c}{Probability of finding $\sigma_{MC}<311.03$  in any quadrant of the simulated maps.}
\tablefoottext{d}{Probability of finding $\sigma_{MC}<359.67$  in any quadrant of the simulated maps.}
\end{table}

\begin{table}
\begin{center}
\caption{Calculated probabilities of finding asymmetries equal to or higher than those found in Planck data in the MC simulations using both Planck and WMAP KQ85 mask considering the $\Lambda$CDM model.}
\label{tbl-prob-mask} 
\begin{tabular}{ccrcrcr}
\hline\hline
 Map & $\sigma_{SEQ}/\sigma_{NWQ}$ & P1 \tablefootmark{a} & $\sigma_{SEQ}/\sigma_{SWQ}$ & P1 & $\sigma_{SEQ}/\sigma_{NEQ}$ & P1 \\
 \hline
 SMICA + WMAP KQ85                 & 6.1  & 0.2\%  &4.7    &0.5\%  & 8.3  & $<$ 0.1\% \\ 
 SMICA +  Planck mask   &5.8   &0.2\%    &3.8    &1\%    & 6.7  & $< $ 0.1\%\\

\hline\hline
 Map & $\sigma_{SEQ}/\sigma_{NWQ}$ & P2 \tablefootmark{b} & $\sigma_{SEQ}/\sigma_{SWQ}$ & P2& $\sigma_{SEQ}/\sigma_{NEQ}$ & P2 \\
 \hline
 SMICA + WMAP KQ85                 & 6.1  & 0.6\%  &4.7    &1.5\%  & 8.3  & $<$ 0.1\% \\ 
 SMICA +  Planck mask   &5.8   & 0.4\%   &3.8    &3.1\%    & 6.7  &  0.1\%\\   \hline
 
 \hline\hline
 Map & $\sigma_{SEQ}/\sigma_{NWQ}$ & P3 \tablefootmark{c} & $\sigma_{SEQ}/\sigma_{SWQ}$ & P3& $\sigma_{SEQ}/\sigma_{NEQ}$ & P3 \\
 \hline
 SMICA + WMAP KQ85                 & 6.1  & 1.8\%     &4.7    &4.8\%  & 8.3  &  0.2\% \\ 
 SMICA +  Planck mask    &5.8   &  1.7\%    &3.8    &11.3\%    & 6.7  & 0.8\%\\   \hline

\end{tabular}
\end{center}
\tablefoottext{a}{Probability of finding the asymmetries in the simulations for exactly same configuration as in the SMICA map.}
\tablefoottext{b}{Probability of finding the asymmetries between the SEQ quadrant  and  any other quadrant in the simulations.}
\tablefoottext{c}{Probability of finding the asymmetries between any pair of quadrants in the simulations.}

\end{table}

\begin{table}
\begin{center}
\caption{ Sigma ratios considering only SWQ, NWQ, and NEQ  for Planck using both Planck and WMAP KQ85 mask. }
\label{tbl_sigma_planck} 
\begin{tabular}{cccc}
\hline\hline
Map&$\sigma_{SWQ}/\sigma_{NEQ}$ &$\sigma_{SWQ}/\sigma_{NWQ}$ &$\sigma_{NEQ}/\sigma_{NWQ}$\\
\hline
SMICA +  Planck mask   & $ 1.74$ &$1.51$ &$0.86$\\
\hline
SMICA +  WMAP KQ85   & $ 1.74$ &$1.28$ &$0.73$\\
\hline\hline 

\end{tabular}
\end{center}
\end{table}

\end{document}